\documentclass[a4paper, amsfonts, amssymb, amsmath, reprint, showkeys, nofootinbib, twoside]{revtex4-1}
\usepackage[english]{babel}
\usepackage[utf8]{inputenc}
\usepackage[colorinlistoftodos, color=green!40, prependcaption]{todonotes}
\usepackage{amsthm}
\usepackage{mathtools}
\usepackage{physics}
\usepackage{xcolor}
\usepackage{graphicx}
\usepackage[left=23mm,right=13mm,top=35mm,columnsep=15pt]{geometry} 
\usepackage{adjustbox}
\usepackage{placeins}
\usepackage[T1]{fontenc}
\usepackage{lipsum}
\usepackage{csquotes}
\usepackage{booktabs}
\usepackage{verbatim} 

\usepackage{tabu}
\usepackage{xcolor}
\usepackage{tabularx}
\usepackage{booktabs}
 
\usepackage[pdftex, pdftitle={Article}, pdfauthor={Author}]{hyperref} 
\begin{document}
\title{Physics-guided descriptors for prediction of structural polymorphs}

\author{Bastien F. Grosso}
    \email[Correspondence email address: ]{bastien.grosso@mat.ethz.ch}
\author{Nicola A. Spaldin}
\author{Aria Mansouri Tehrani}
    \affiliation{Materials Theory, ETH Zürich, Wolfgang-Pauli-Strasse 27, 8093 Zürich, Switzerland}
    
\date{\today} 


\begin{abstract}
We develop a method combining machine learning (ML) and density functional theory (DFT) to predict low-energy polymorphs by introducing physics-guided descriptors based on structural distortion modes. 
We systematically generate crystal structures utilizing the distortion modes and compute their energies with single-point DFT calculations. We then train a ML model to identify low-energy configurations on the material's high-dimensional potential energy surface. Here, we use BiFeO$_3$ as a case study and explore its phase space by tuning the amplitudes of linear combinations of a finite set of distinct distortion modes. Our procedure is validated by rediscovering several known metastable phases of BiFeO$_3$ with complex crystal structures, and its efficiency is proved by identifying 21 new low-energy polymorphs. This approach proposes a new avenue toward accelerating the prediction of low-energy polymorphs in solid-state materials.

\end{abstract}

\maketitle

\section{Introduction}

Computational materials science has undergone a recent paradigm shift with the advent of data-driven methods such as machine learning (ML). These techniques are now considered a standard tool, along with widely used methods such as density functional theory (DFT), molecular dynamics, or Monte-Carlo simulations. 

One central task for effectively applying machine learning algorithms to materials science problems is developing appropriate descriptors. Descriptors are vector-based numerical representations that should uniquely define the material and are mainly based on compositional or structural features or a mixture of both. In addition, these descriptors should provide meaningful connections to the physics of the materials by establishing a unique and invariant ``barcode'' for each one of them.

The advantage of descriptors based only on the chemical composition such as atomic number, covalent radius, number of valence electrons, etc., \cite{Zhuo/MansouriTehrani:2018} is that no prior knowledge of the system is required. However, while this approach has been successful in predicting different properties such as band gap, hardness, and thermodynamic stability,\cite{deJong/Chen/Notestin:2016,Li/Jacobs/Morgan:2018,Zhang/MansouriTehrani/Oliynyk:2020,Wang/Kauwe/Murdock:2021} it is not capable of distinguishing between different structures of the same composition. Therefore, it is crucial to incorporate the influence of structural features to describe phenomena such as metal-insulator transitions or ferroelectricity correctly.\cite{Balachandran/Rondinelli:2013,He/Li/Liu:2021,Frey/Grosso:2022} In contrast, descriptors based on structural features, such as the Smooth Overlap of Atomic Positions (SOAP),\cite{Bartok/De/Poelking:2017} Minimum bounding ellipsoid (MBE),\cite{Cumby/Attfield:2017} ordered eigenvalues of the Coulomb matrix \cite{Rupp/Tkatchenko/Muller:2012}, or universal fragment descriptors \cite{Isayev/Oses/Toher:2017} have been developed in order to capture the local environment of the atoms and encode structural features. 

A natural approach to representing a crystal structure is to use the concept of irreducible representations (irreps), in which the distortions from a higher symmetry reference structure are decomposed into a set of normal modes that describe the transformation of the reference structure into the considered one.\cite{Stokes/Hatch/Campbell,Campbell/Stokes/Tanner:2006}. This approach has been used to understand and explain various phenomena such as thermopower anisotropy, negative thermal expansion, proper and improper ferroelectricity, and antiferroelectricity.\cite{Puggioni/Rondinelli:2014,SennBombardi/Murray:2015,Nowadnick/Fennie:2016,Bostroem/Senn/Goodwin:2018,Shapovalov/Stengel:2021} Furthermore, in the context of phase transitions in transition-metal compounds, the distortion modes have been used as features in statistical analysis of the correlation between distortions and functionalities.\cite{Balachandran/Rondinelli:2013,Balachandran/Benedeck/Rondinelli:2015,Wagner/Puggioni/Rondinelli:2018} More recently, polyhedral distortions have been used as descriptors to explain trends in behaviors across perovskites. \cite{Morita/Davies/Butler:2022}

In this work, we explore the idea of using the distortion modes as descriptors in ML. Our method utilizes distortion modes to explore the Born-Oppenheimer potential energy surface (PES) and identify local minima corresponding to metastable phases (polymorphs). While the method that we introduce remains applicable to any crystalline material, here we focus on perovskites for which the distortion modes are simply obtained as a decomposition of the high symmetry prototype-perovskite parent structure with $Pm\bar{3}m$ symmetry. We apply the method to multiferroic bismuth ferrite, BiFeO$_3$ (BFO), which exhibits simultaneous ferroelectricity and magnetic ordering and has a rich structural landscape composed of many local minima with interesting technological properties. \cite{Dieguez/Gonzalez-Vazquez/Wojdel:2011,Grosso/Spaldin:2021}  

The ground-state structure of BFO (\emph{R3c}) has a 10-atom unit cell and is reached from the $Pm\bar{3}m$ prototype by a combination of polar displacement of the Bi-cation (mode at $\Gamma$) and anti-phase rotation of consecutive oxygen octahedra in all three cartesian directions (mode at $R$).\cite{Kubel/Schmid:1990} Several metastable phases have been established computationally, many with larger unit cells.\cite{Grosso/Spaldin:2021} Furthermore, the experimental stabilization of many new phases of BFO \cite{Nordlander/Maillard:2020,Mundy/Grosso/Heikes:2022,Caretta/Shao/Yu:2022} motivates our choice of BFO both to test the predictive power of the method and to identify new metastable phases for future experimental study. 

The remainder of this paper is organized as follows. In Sec. \ref{sec:methodology}, we present the computational details of our calculations, explain how we set the distortion modes, and present an overview of our workflow. In Sec. \ref{sec:results} we present our results following the workflow of our method: First, we train our machine learning model on a set of energies and structures that we create using DFT (Sec. \ref{sec:round1}). Then, we use the developed model to search for structural polymorphs by predicting the energies of a large number of structures and selecting the most promising structures for further analysis using DFT (\ref{sec:round2}). Finally, we conclude our study and give possible extensions to our work in Sec. \ref{sec:conclusion}. The codes used in this work are provided in the open-source \textcolor{red}{GitHub} repository.


\begin{center}\
\begin{figure*}[th]\
\includegraphics[width=1\linewidth]{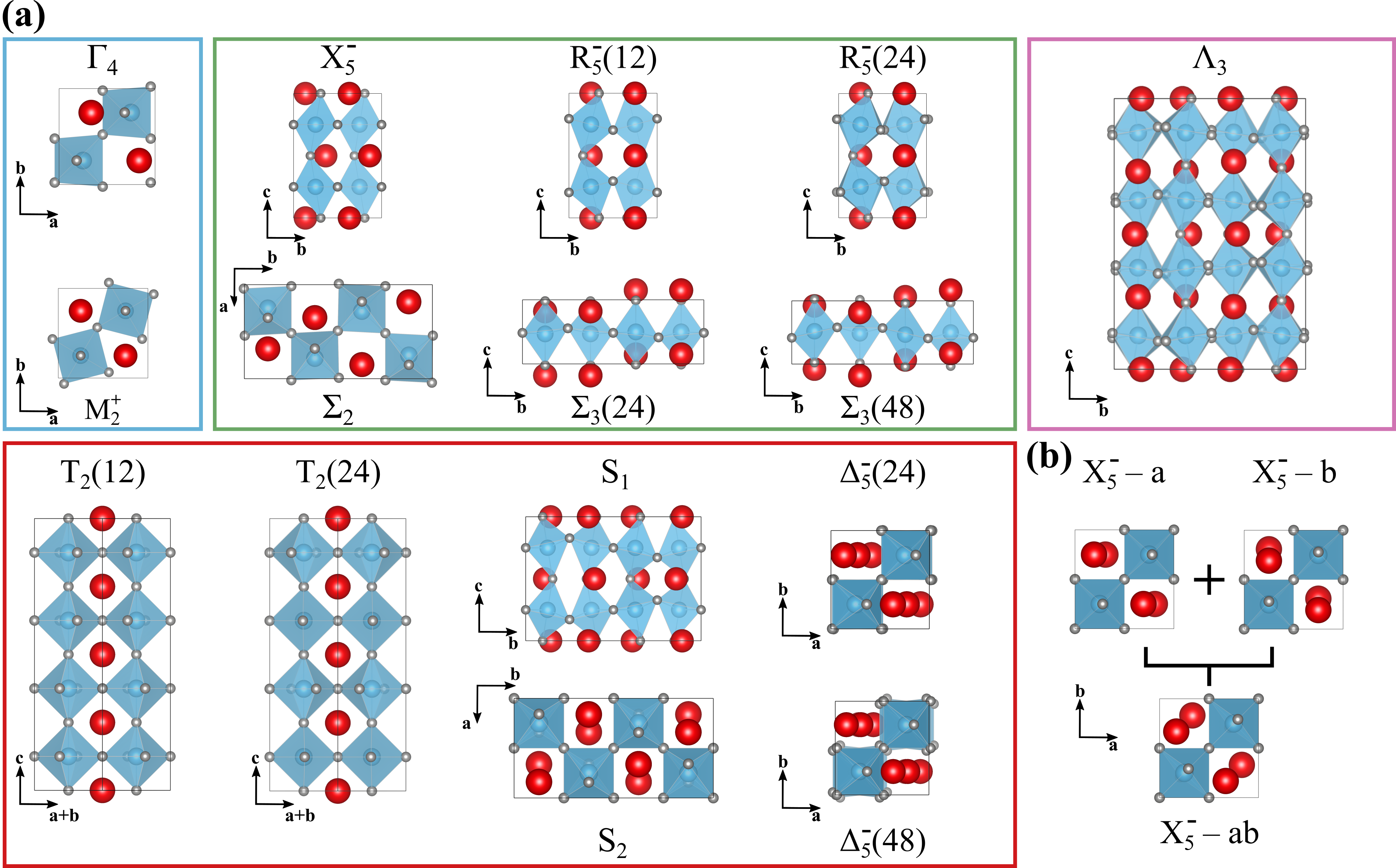}
\caption{(a) Structures of the different distortion modes used as descriptors. The structures are gathered according to the number of atoms each primitive unit cell contains and contain 10, 20, 40, and 80 atoms per unit cell for the blue, green, red, and pink boxes, respectively. Note that we distinguish between distortion modes with the same symmetry by indicating the corresponding symmetry index in parentheses. (b) The structure of a distortion mode along orthogonal directions illustrates how we create extra descriptors with different orientations. Since the structures corresponding to the distortions imposed by the $X_5^-$ mode along $\mathbf{a}$ and $\mathbf{b}$ are identical, we include in our set of building blocks the symmetrically distinct structures obtained from $X_5^- - a$ the $X_5^- - ab$.  }
\label{Fig1}
\end{figure*}\
\end{center}\

\section{Methodology}\label{sec:methodology}
\subsection{Computational details}\label{sub:compdetails}
The DFT calculations are performed using VASP \cite{Vasp1:1993,Vasp2:1994,Vasp3:1996,Vasp4:1996} with the PAW method \cite{Blochl:1994,Kresse/Joubert:1999} and explicit treatment of the following valence electrons: $6s^26p^3$ for Bi, $3d^74s^1$ for Fe and $2s^22p^4$ for O. A $8 \times 8 \times 6$ k-point $\Gamma$-centered Monkhorst-Pack mesh \cite{Monkhorst/Pack:1976} is used to sample the Brillouin zone of a $20$-atom unit cell, and an energy cutoff of $700$ eV for the plane-wave basis is chosen. For the xcs functional we use PBEsol \cite{PBEsol:2008} with an effective Hubbard-like correction, $U_{eff} = 4$ eV, for the Fe $d$ orbitals according to Dudarev's approach \cite{Dudarev/Botton/Savrasov:1998}. G-type antiferromagnetism is adopted for all the calculations.

The machine learning model is created based on a Support Vector Machine Regression (SVR) method \cite{Chang/Lin:2011,SVR_Drucker:1997} as implemented in the Scikit-learn \cite{Scikit_learn:2011} library. We use a SVR model with a radial basis function (RBF) as the kernel function. Using the approach presented in Sec. \ref{sec:round1}, we generate $9,569$ structures and energies using DFT (see Sec. and split the data into a training set ($80\%$, $7,655$ data points) and a test set ($20\%$, $1,914$ data points). We further perform a five-fold cross-validation scheme to optimize the cost constant ($C$) and the RBF free parameter ($\gamma$) to be $C = 10$ and $\gamma = 1$ for the best performance. We finally use these parameters to evaluate the model on the test set.

\subsection{Distortion modes as building blocks of structures} \label{sub:blocks}
\subsubsection{General considerations}
When a structure experiences a symmetry lowering, part of its symmetry elements are lost, and the remaining symmetry operations constitute a subgroup of the initial parent space group. If there exists a group-subgroup relation between the structures, one can decompose the distorted one into symmetry-adapted modes of the parent structure that encode patterns of displacements of the atoms. For example, the perfect perovskite structure with space group $Pm\bar{3}m$ can be considered as the higher symmetry parent structure of all lower symmetry structures that are distorted versions of the $Pm\bar{3}m$ phase. It becomes then clear that different subgroups of $Pm\bar{3}m$ can share common symmetry modes. Nevertheless, each structure is characterized by a unique combination of symmetry modes with given relative amplitudes. 

As each mode consists of a displacement pattern of the atoms, the combination of different modes can either lower or increase the structure's total energy compared to the undistorted perovskite structure. The increase in energy can occur, for example, in the case of the interatomic distances between anions and cations becoming too small. While chemical intuition could, in principle, allow one to create simple subgroups of distortion modes, resulting in (meta-)stable structures, the number of possibilities and the complexity of many distortions are prohibitively large, and a more systematic approach, which we propose here, is required.

\subsubsection{The model case of BFO}

Our study starts with selecting the distortion modes that will form our descriptors. To select appropriate distortion modes, we extract from the phases previously reported in Refs. \cite{Dieguez/Gonzalez-Vazquez/Wojdel:2011}, and \cite{Grosso/Spaldin:2021} all the different modes that are contained in unit cells of 80 atoms or fewer using ISODISTORT.\cite{Stokes/Hatch/Campbell,Campbell/Stokes/Tanner:2006} This yields the 15 modes shown in Fig. \ref{Fig1}a. Since certain modes result in displacements that can arbitrarily be applied along different directions, we include 6 additional building block structures obtained by linear combinations of the non-rotational distortion modes along the directions defined by the lattice vectors. We show an example in Fig. \ref{Fig1}b with the $X^-_5$ mode, where we take the linear combination of the structures obtained by this mode applied along two different directions ($\textbf{a}$ and $\textbf{b}$) and obtain a structure where the distortions are along $\textbf{ab}$. While the structures with the distortions along $\textbf{a}$ and \textbf{b} are identical, the new structure with the distortions along the diagonal of the unit cell has a different symmetry and can then be considered as a building block, keeping the uniqueness of the descriptors. We then obtained 21 building blocks constructed by distorting a perfect cubic supercell according to 15 different modes and six of their spatial variations.

\begin{center}
\begin{figure}[ht]
\includegraphics[width=\columnwidth]{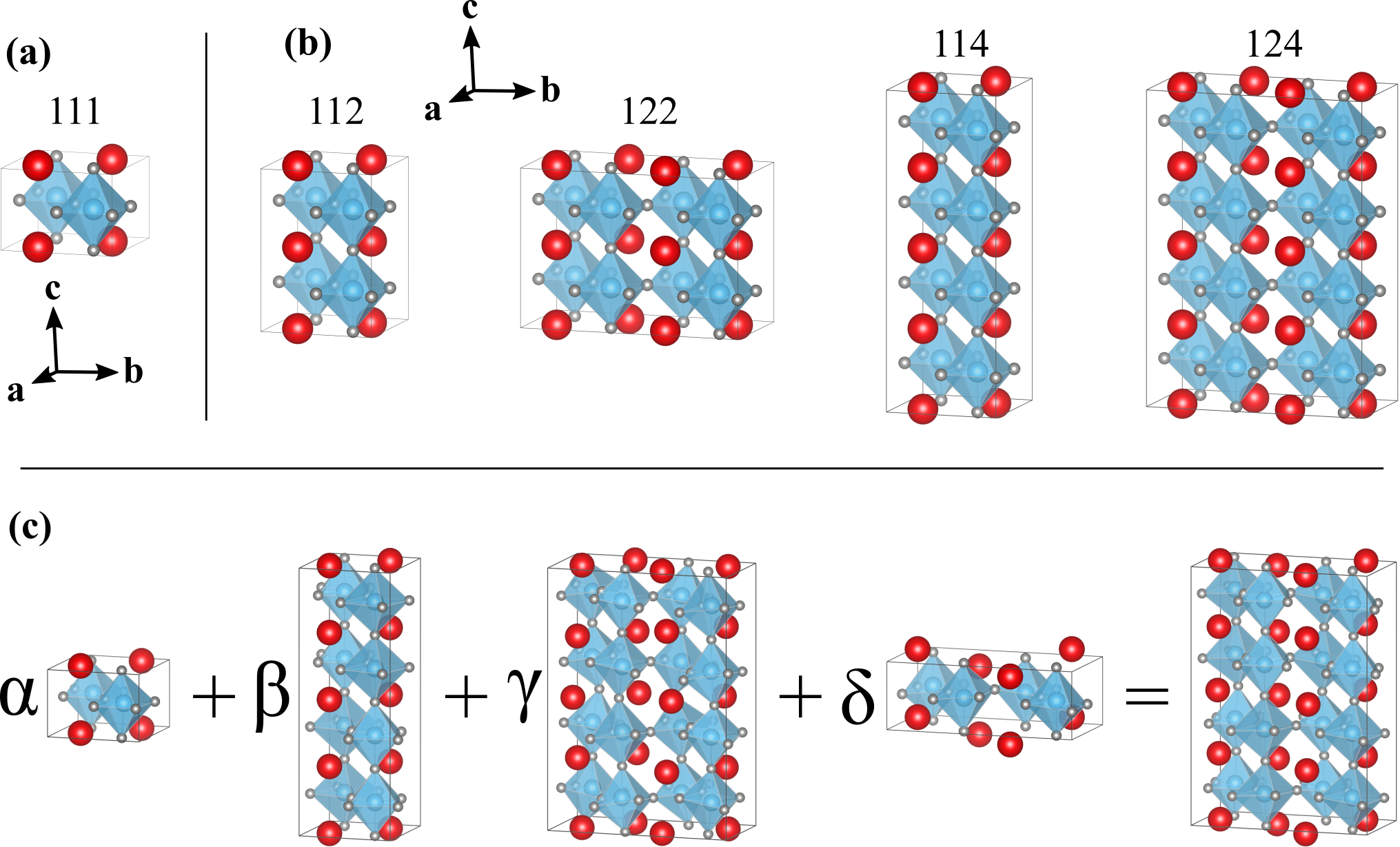}
\caption{(a) Basic cubic unit cell labeled $111$. (b) Possible supercells. Each supercell is constructed from multiples of the cubic unit cell structure in (a) along each cartesian direction and labeled accordingly. (c) Example of a combination of four modes to build a structure. All structures of the modes are scaled to the size of the structure of the third mode (which has the largest supercell), and for each mode, we create a distortion vector capturing the displacement from the $124$ structure displayed in (b) and the considered mode. Each vector is then multiplied by the corresponding amplitude factor ($\alpha, \beta, \gamma$ or $\delta$), and the vectors are summed to give the final structure.}
\label{Fig2}
\end{figure}
\end{center}

Using symmetry considerations, we reduce the unit cell size of the distorted structure to its minimal possible size. We then fix its volume to match that of a supercell of the $Pm\bar{3}m$ structure (Fig. \ref{Fig2}a) coinciding with the distorted structure. This constrains the structure to four possible unit cell sizes displayed in Fig. \ref{Fig2}b, all of which accommodate G-type antiferromagnetic order. Furthermore, we normalize all the modes in such a way that the sum of the displacements multiplied by $\sqrt{V_p/V_s}$ is equal to 1 \AA, equivalent to a mode amplitude of 1, where $V_p$ and $V_s$ are the primitive and supercell volumes respectively. This allows us to describe evenly all modes, no matter the size of their unit cell.

Finally, a structure can be created by displacing the atoms in the parent structure by a sum of displacements resulting from each normalized mode multiplied by a scalar amplitude, as represented in Fig. \ref{Fig2}c. The displacements are computed in supercells corresponding to the largest structure present in the set of modes selected (see Fig. \ref{Fig2}b)

\subsection{Workflow}

\begin{center}
\begin{figure}[ht]
\includegraphics[width=\columnwidth]{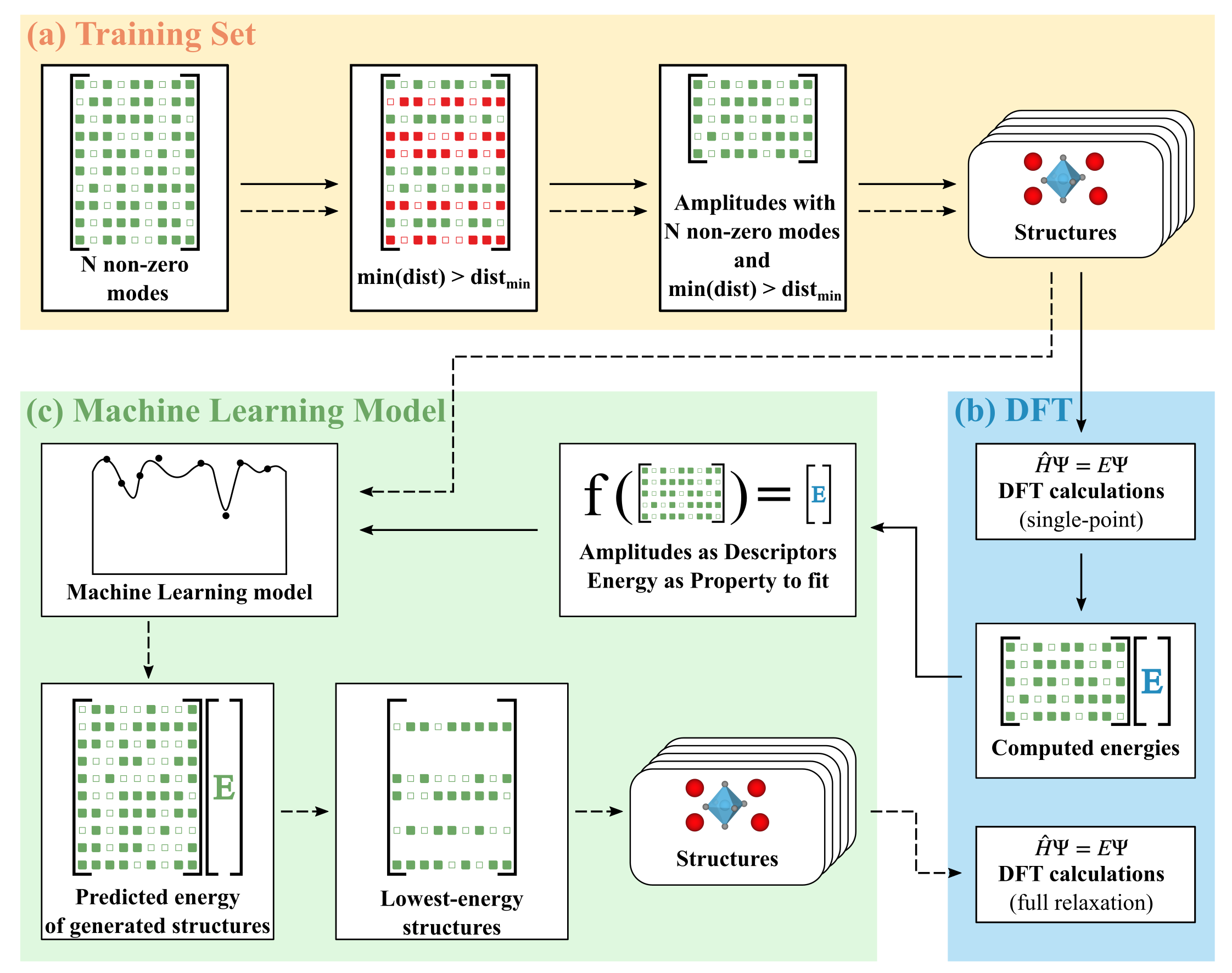}
\caption{Overview of our DFT-ML procedure. The first round is indicated by solid arrows and the second by dashed ones. (a) (yellow) Construction of the training set. For a given value of $N$, we start from a list of $N$ non-zero random amplitudes (first panel), where each row of the list corresponds to a structure and each column to the amplitude of a given distortion mode. Non-zero amplitudes are represented by full squares and zero amplitudes by empty squares. From the list, we discard (red) the structures in which two atoms are closer than the minimum distance allowed (dist$_{min}$) (second and third panels). Finally, we generate the input files corresponding to the structures (fourth panel) that we retain for the DFT section. (b) (blue) DFT calculations for electronic relaxation (b - top) or electronic, ionic, and volumetric relaxation (b - bottom). We compute the energy with a single-point DFT calculation for each line of the list. (c) Using ML, we use the list of amplitudes gathered in the previous step as descriptors and fit the corresponding energies. We then obtain a model that can predict the energy of a given combination of amplitudes. Dashed arrows represent the second round. Following the same steps as described in the first round, we (a) generate structures with different values of $N$. Then, we use the ML model created to predict the energy of those structures, select the lowest energy ones and create the corresponding structures (c). Finally, we use these structures as starting configurations and fully relax the ions, lattice parameters, and electrons (b).}
\label{Fig3}
\end{figure}
\end{center}

In Fig. \ref{Fig3}, we show our combined DFT and ML approach, which proceeds in two rounds. In the first round, the model is trained by building a training set on distortion modes and computing the energies of the structures using DFT single-point calculations. In this stage, only the electrons are relaxed while the ions are kept fixed to determine the energies as a function of the modes' amplitudes. In the second round, using the same methodology, we generate a large number of structures and utilize our ML model, instead of the more expensive DFT calculations, to predict their energies. Finally, we perform DFT ionic relaxation for a selected number of low-energy structures.
This approach allows us to start the DFT relaxation on initial structures predicted to be low energy by our ML model and thus are likely to be close to a local minimum of the potential energy surface.

\section{Results and Discussions}\label{sec:results}

\subsection{First round: building the ML model}\label{sec:round1}
The first round starts with the generation of the training set. We consider the potential energy surface of BFO in a $21$-dimensional space where each dimension is given by a distortion mode, and each coordinate in the space is given by the amplitude of the modes. To span the energy surface unbiasedly, we randomly generate the amplitudes of the modes in the interval $[-1.2;1.2]$. This approach generates amplitudes corresponding to the sum of the atomic displacements from the parent structure. The choice of the $[-1.2;1.2]$ range, even though arbitrary, is set to capture a wide range of the potential energy for each mode. Furthermore, to account for the physics of each mode individually and to capture any favorable or unfavorable couplings up to the seventh order between different modes, we include structures generated by including up to $7$ modes with amplitudes different than zero.

Our procedure starts (Fig. \ref{Fig3}a) by choosing an order of coupling $N$, equal to the number of non-zero amplitudes, and generating a matrix of $L$ lines and $21$ random numbers in the interval $[-1.2;1.2]$ for each line, where $L$ is the number of potential structures. We then obtain a matrix of dimension $L \times 21$. Secondly, we randomly set amplitude entries per line to zero ($21$-$N$). We then discard non-physical structures, determined by computing the minimum distance between pairs of atoms in each structure and removing the structures for which any minimum distance is too small. Based on the distribution of the Fe--O bond lengths in all structures containing these ions in the ICSD database \cite{Oliynk_unpublished,ICSD_Levin}, we keep only structures with Fe--O spacing larger than $1.8$ \AA. This procedure is repeated for each value of $N$ considered and results in a set of structures ready for the following step. Note that all structures' volumes are fixed to integer multiples of the cubic undistorted cell.

We then take all the created structures and compute their energies with a single-point DFT calculation in which only one electronic loop is computed. As a result, we obtain the energy for each structure described by a row of amplitudes in the matrix (Fig. \ref{Fig3}b, top). 

\begin{center}
\begin{figure}[ht]
\includegraphics[width=\columnwidth]{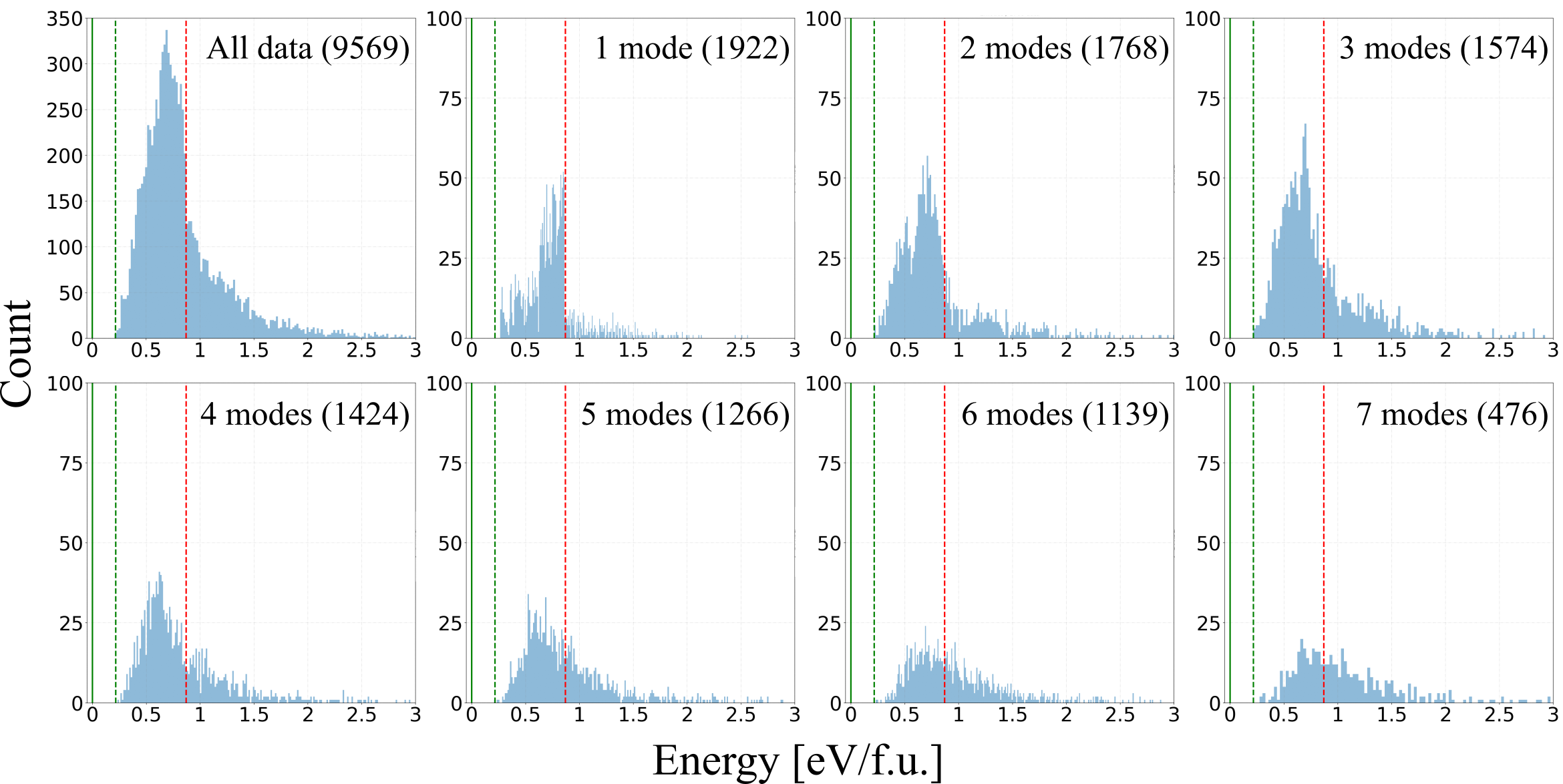}
\caption{Histograms of the energies of the structures in the training set. The whole training set generated combines $9569$ structures (top-left) distributed in structures combining, from top-left to bottom-right, only $1$,$2$,$3$,$4$,$5$,$6$ and $7$ modes, respectively. The number of structures included for each mode is indicated next to the corresponding legend in parenthesis. The continuous green vertical line represents the reference energy of the $R3c$ ground state. The dashed green vertical line shows the energy of the constrained $R3c$ structure where the $R3c$ coordinates have been constrained to the cubic parent structure (volume and angles), and the dashed red vertical line displays the energy of the cubic parent structure (amplitudes of all the modes set to zero). The bins have a width of $10$ meV in all plots except the $7$ modes one, where the bin size is $30$ meV for better visualization.}
\label{Fig4}
\end{figure}
\end{center}

We present in Fig. \ref{Fig4} the energies of the different structures included in the training set as well as the number of structures per mode. We see that most of the structures included in the training set have energy lying between that of the $R3c$ phase constrained to the cubic parent structure volume and angles and that of the cubic parent structure. Furthermore, when the number of modes present in the structure increases, the energy tends also to increase. This is a result of our choice of random amplitudes, for which additional distortions reduce the likely distance between ions and consequently increase the energy.

\begin{center}
\begin{figure}[h!]
\includegraphics[width=\linewidth]{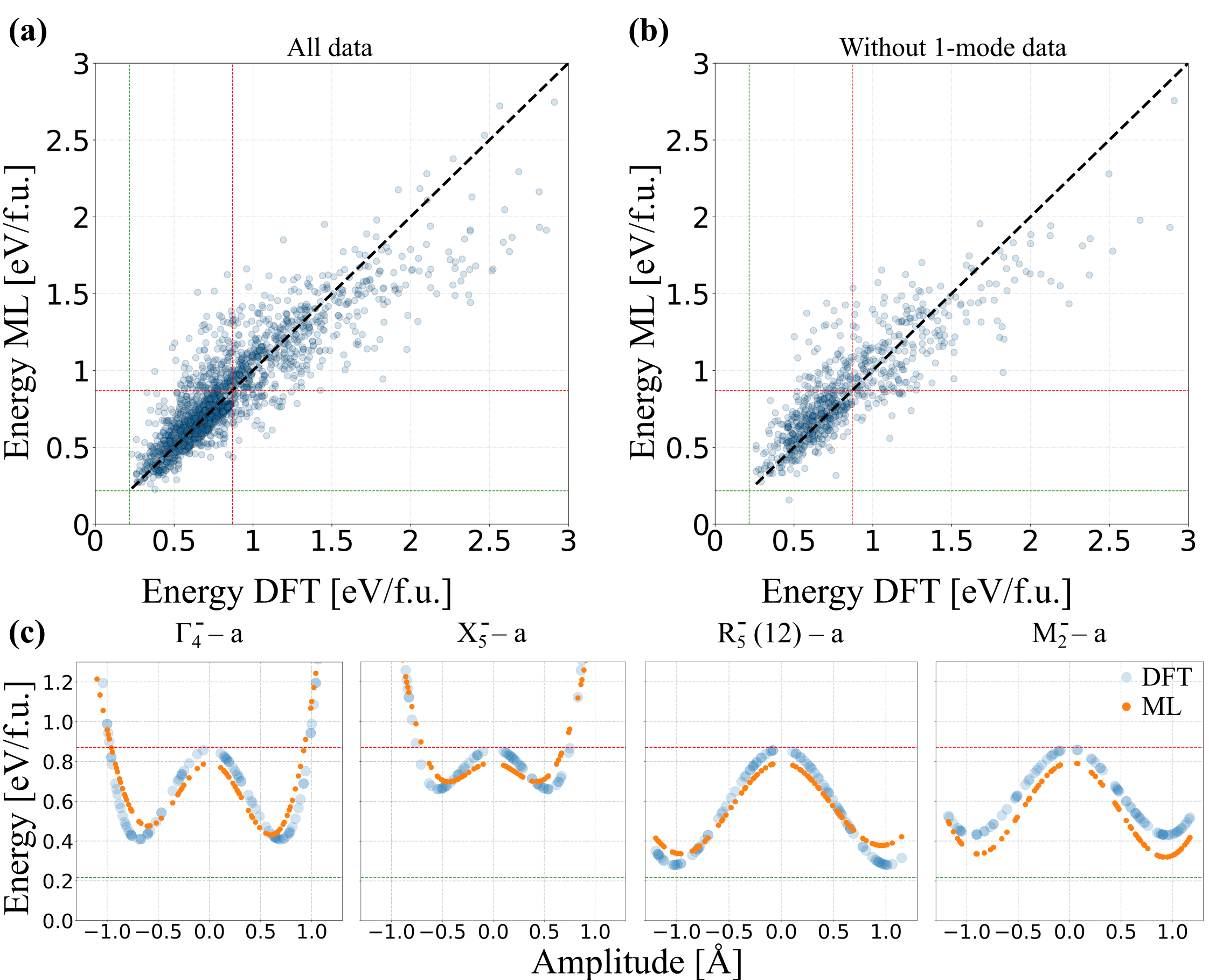}
\caption{Evaluation of the prediction accuracy of our ML models. (a) Regression plot of predicted (ML) energies versus computed (DFT) energies on the test set. (b) Predicted energy versus computed energy when all the structures containing only one mode were removed from the training set. Each blue dot represents the energy of a structure. The black dashed line represents the position of perfect predictions, where the energy predicted by the ML model perfectly agrees with the DFT energy. The red and green lines are the energy of the cubic and $R3c$-like structures, respectively. (c) Computed energies (blue) compared to the energies predicted with the model fitted on the data without the 1-mode structures (orange) for four often discussed modes. The red and green lines have the same meaning as in (a) and (b).}
\label{Fig5}
\end{figure}
\end{center}

The last step of the first round (see Fig. \ref{Fig3}c) consists of using the training set to fit the energies as a function of the modes' amplitudes. We use a Support Vector Regression (SVR) algorithm, which outperformed other machine learning algorithms that we tried, such as Random Forest and Extreme Gradient Boost (XGB), to construct a model evaluated on the test set. Fig. \ref{Fig5}a illustrates the reasonable agreement between the machine learning predicted total energies and DFT calculations. The coefficient of determination ($r^2$) is around 0.77, and the root-mean-square error (RMSE) is determined to be 0.2 eV/f.u.. It is worth noting that within the range of energy contained by the constrained $R3c$ and parent cubic structures (lower left of the plot), the points are less spread around the diagonal line (ML = DFT), than they are in the upper right of the plot. This can be attributed to the lower representation of high-energy structures in the training set.

Next, we use the permutation feature importance technique to evaluate which features (distortion modes) significantly contribute to our SVR model prediction of the test set. In this technique, we randomly shuffle the value of each feature 30 times and evaluate the variation in the $r^2$ metric. A feature is deemed important if the shuffling leads to substantial changes in the model's error. Fig. \ref{FigS1} in the supporting information shows the permutation importance of each feature based on changes in $r^2$ sorted from highest to lowest. Based on this analysis, the $X_5^{-}$ mode along $\textbf{ab}$ is the most important feature, and, as expected, the $\Gamma_4^{-}$ modes are also significant. However, surprisingly, modes corresponding to 40-atom supercells, such as $T_2(24)$ or $\Delta_5^-(24)$, are also crucial for the model to predict the energy of various structures, demonstrating the importance of including these less often discussed modes.  

We further evaluate the extrapolative power of our model by removing all the structures that contain only one distortion mode ($1922$ points) from the training set before predicting their energies. Fig. \ref{Fig5}b show the predicted energy values using the SVR model against the DFT energies, constructed on the training set without single distortion modes. As expected, the $r^2$ evaluated on the test set is now lower ($\approx\,74\%$) with RMSE = 0.21\,eV/f.u. compared to the full test set.

Next, utilizing the model built without the single distortion modes, we test whether our machine learning model accurately reproduces the energy variation of the distortion modes with respect to the amplitude by comparing the ML energies to their corresponding DFT energies. Our results are shown in Fig. \ref{Fig5}c. We see that the SVR model correctly predicts the signature double-well behavior for the $\Gamma_4^{-}-a$, $X_5^{-}-a$, $R_5^{-}(12)-a$, and $M_2^{-}-a$  modes, as well as determining the high energy barrier between the two minima for $R_5^{-}(12)-a$ and $M_2^{-}-a$. These curves are also plotted for the rest of the individual distortion modes in Fig. \ref{FigS2} of the supporting information, where we see that good agreement between DFT and ML predictions is obtained for most of them. Thus, despite showing imperfect statistical metrics on the test set, the ML model successfully captures the structural energetics of BFO.

\subsection{Second round: wide exploration of the PES}\label{sec:round2}

In the second round, we use the constructed model to predict the energies of a large number of structures built with the random composition of $1$ to $7$ modes combined. First, we construct $457,081$ structures in the same way as described in section \ref{sec:round1} and as shown schematically in Fig. \ref{Fig3}a and predict their energies. Using the model, we select the structures within $250$ meV/f.u. of the ground state and classify them into subsets according to their distortion modes. Finally, we select the lowest energy structure of each subset. As a result, we obtain $290$ structures, all with different combinations of the modes and containing 20, 40, or 80 atoms, that we then fully relax with DFT.  

\setlength{\tabcolsep}{10pt}
\renewcommand{\arraystretch}{1.0}

\begin{table*}[h!]
  \centering
\resizebox{1\textwidth}{!}{%
  \begin{tabular}{ l r r r c c}
  \toprule[1.5pt]
   &
  \multicolumn{4}{c}{\textbf{Lattice parameters}} &
  \multicolumn{1}{c}{\textbf{Energies}}
  \\

  \cmidrule[0.7pt](lr){2-5}
  \cmidrule[0.7pt](lr){6-6}

 \textbf{Symmetry}& a [\r{A}] & b [\r{A}] & c [\r{A}]  & Angles [deg.] & Energy [meV/f.u.]  \\
   \cmidrule[1.4pt](lr){1-6}
{P1} & $5.54$ & $11.14$ & $15.55$ & $\alpha = 90.020, \beta = 90.087, \gamma = 90.023$ & $48$ \\ 
& $5.55$ & $11.17$ & $7.76$ & $\alpha = 90.175, \beta = 90.012$ & $54$ \\ 
& $5.52$ & $5.54$ & $15.84$ & $\alpha = 90.492, \beta = 90.233, \gamma = 90.061$ & $57$ \\ 
{P$2_1$} & $5.52$ & $11.04$ & $15.68$ & $\beta = 90.272$ & $44$ \\ 
& $5.55$ & $11.02$ & $15.62$ &  $\beta = 90.085$ & $54$ \\ 
{Pm} & $5.52$ & $15.67$ & $11.10$ & $\beta = 90.128$ & $45$ \\ 
& $5.53$ & $11.09$ & $7.78$ & $\beta = 90.006$ & $56$ \\ 
{Cc} & $7.79$ & $7.81$ & $15.92$ & $\beta = 90.623$ & $68$ \\ 
{P$2_1$/m} & $5.52$ & $11.11$ & $15.66$ & $\beta = 90.269$ & $46$ \\ 
{P$2_1$/c} & $5.54$ & $11.15$ & $7.76$ & $\beta = 90.301$ & $64$ \\ 
& $5.49$ &  $11.15$ & $15.67$ & $\beta = 90.707$ & $68$ \\ 
{Pmc$2_1$} & $5.55$ & $11.00$ &  $7.79$ &  & $41$ \\ 
& $5.53$ & $11.10$ & $15.65$ &  & $42$ \\ 
& $5.52$ & $11.15$ & $7.77$ & & $69$ \\ 
& $5.50$ & $5.56$  &$15.53$ &  & $96$ \\ 
{Pca$2_1$} & $5.48$ & $5.57$ & $15.75$ &  & $23$ \\ 
{Pmn$2_1$} & $5.53$ & $5.59$ & $7.85$ & & $88$ \\
{Cmc$2_1$} & $5.62$& $11.15$ & $7.81$  &  & $96$ \\ 
{Ima$2$} & $5.58$ & $5.59$ & $7.75$ & & $77$ \\ 
& $5.58$ & $5.59$ &$15.49$ &  & $82$ \\ 
{Pbca} & $5.57$ & $11.04$ &$15.53$ &   & $55$ \\ 
  \bottomrule[1.5pt]
  \end{tabular}
  }
  \caption{New low-energy phases identified in this work. We give the symmetry (space group), lattice vectors, angle (indicated if different than $90^\circ$) and energy relative to the fully relaxed $R3c$ ground state. The symmetries are determined using Pymatgen \cite{PingOng/DavidsonRichards/Jain:2013} with a tolerance of $10^{-3}$ \AA.}
  \label{tab_ML_phases}
\end{table*}


Focusing only on the structures displaying an energy of $100$ meV/f.u. or less above the $R3c$ ground state after full relaxation, we obtain $29$ different structures among which $8$ are rediscovered phases: $4$ phases are the common $R3c$, $Pnma$, $Pbam$ and $Pmc2_1$ \cite{Yang/Ren/Stengel:2012,Kozlenko/Belik/Belushkin:2011,Dieguez/Gonzalez-Vazquez/Wojdel:2011} and $4$ were reported in Ref. \cite{Grosso/Spaldin:2021} and labeled $Pc(1)$, $Pnma(1)$, $Cmc2_1(2)$ and $P2_1/c$. We report the $21$ new phases in Tab. \ref{tab_ML_phases}. We observe that the majority of the phases found have $80$ atoms per unit cell and have an energy around $40$-$50$ meV/f.u. above the ground state. While this is the largest unit cell size allowed by our way of building the structures (see Fig. \ref{Fig2}b), it suggests that even larger unit cell could display stable phases. 

\begin{center}
\begin{figure}[h!]
\includegraphics[width=\linewidth]{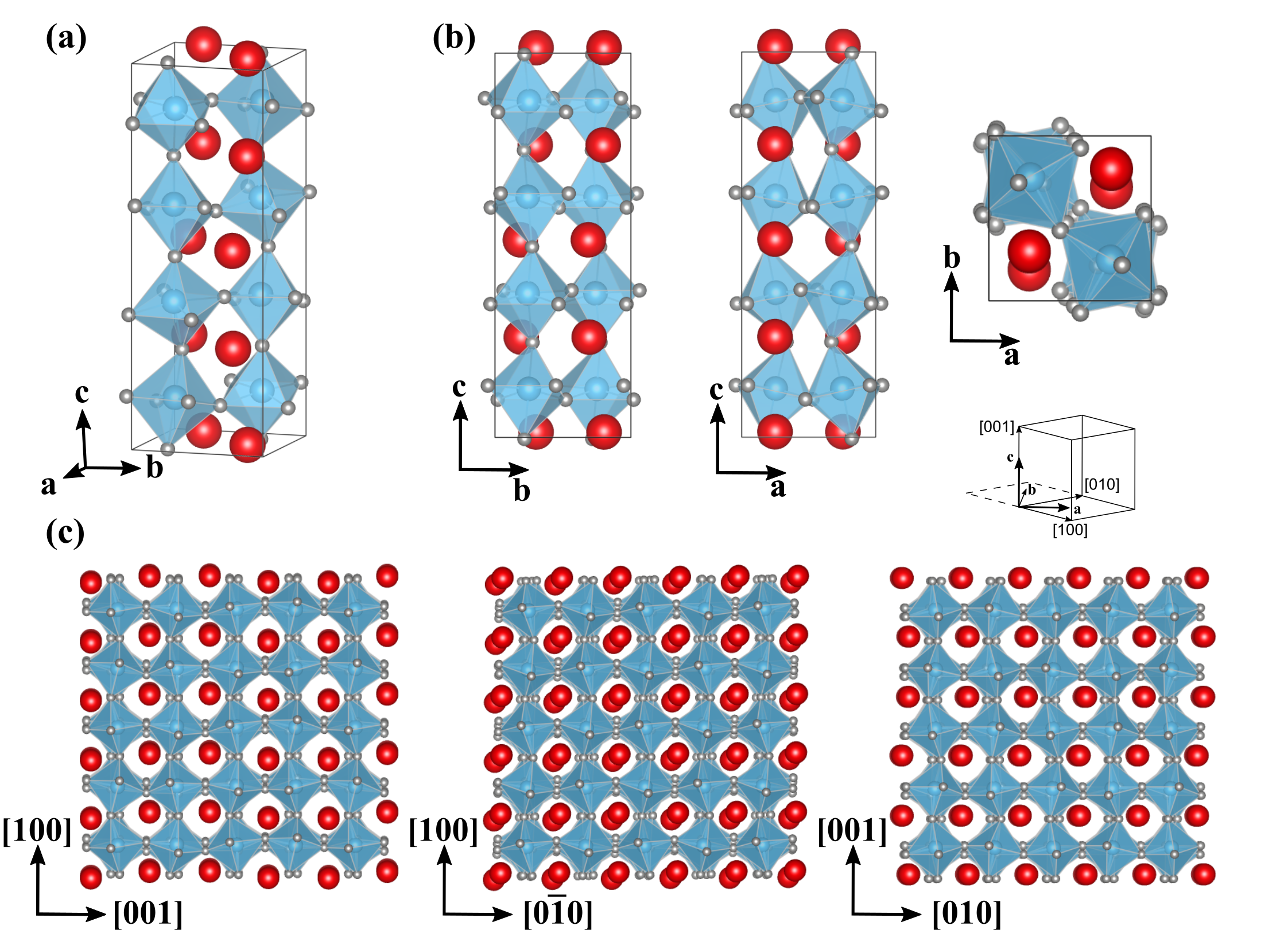}
\caption{Crystal structure of the $Pca2_1$ phase. (a) Full unit-cell. (b) b-c, a-c and a-b projections of the unit-cell. (c) $Pca2_1$ supercells imaged along the three pseudo-cubic directions.}
\label{Fig6}
\end{figure}
\end{center}

While a full analysis of all these phases are out of the scope of this work, we provide information about their structures in Tab. \ref{tab_ML_phases} and focus on the analysis of the lowest-energy phase found with $Pca2_1$ symmetry and present its crystal structure in Fig. \ref{Fig6}. Note that all the crystal structures are available at \textcolor{red}{GitHub}. Its unit cell, as shown in Fig. \ref{Fig6}a, is similar to the $Pc (1)$ phase referenced in \cite{Grosso/Spaldin:2021}, and it has the same energy difference with respect to the ground state ($23$ meV/f.u.). While in the $Pc(1)$ structure, the Bi cations are displaced perpendicular to the long axis, with three cations moving in one direction and one in the opposite direction, in the $Pca2_1$ structure, two Bi cations move in one direction and the next two in the opposite direction (see Fig. \ref{Fig6}b). The $Pca2_1$ structure also has a different rotation pattern with $a^{\bar{\alpha}\beta}a^{\bar{\alpha}\beta}c^{\bar{\alpha}\beta\bar{\gamma}\bar{\delta}}$ rotations, as opposed to the $a^{\bar{\alpha}\beta}a^{\bar{\alpha}\beta}c^{\bar{\alpha}\beta\bar{\gamma}\delta}$ rotations in $Pc(1)$ (notation adopted from Ref. \cite{Grosso/Spaldin:2021}). Considering the isotropic Born effective charges of 4.86 [e] for Bi, 3.99 [e] for Fe, and $-$2.95 [e] for O in units of the electronic charge magnitude, and multiplying the atom displacements with respect to the cubic parent structure by the corresponding isotropic charge,\cite{Grosso/Spaldin:2021} we evaluate the spontaneous polarisation to be around $54$ $\mu$C/cm$^2$ along the c axis (long axis). In Fig. \ref{Fig6}c we present views of the structures along the pseudo-cubic orientations and the structures that could be experimentally observed using high-resolution transmission electron microscopy. In particular, we see that for the view down the c-axis in-plane, a ``$2$up-$2$down'' displacement pattern of the Bi cations appears along the $[100]$ direction, reminiscent of the recently discovered antiferroelectric $Pnma$ phase. \cite{Mundy/Grosso/Heikes:2022}

\section{Summary and Conclusions}\label{sec:conclusion}
We introduced a new DFT-ML approach to efficiently explore the energy landscape of complex solid-state materials by implementing distortion modes as descriptors. Our approach was successfully implemented within the BiFeO$_3$ phase space, rapidly rediscovering low-energy crystal structures that had previously been identified using various computational and experimental methods. In addition, we predicted 21 new low-energy polymorphs of BFO (less than 100 meV/f.u. above the ground state), including one of the lowest-energy polymorphs of BFO reported, with the $Pca2_1$ symmetry and showing a large polarization. Our predictions (all crystallographic informations of the predicitons can be accessed through the GitHub link) further highlight the rich phase-space of BFO and hopefully motivates additional experimental work to synthesize these predictions. 

 Our approach of utilizing distortion modes as building blocks provides a facile way to generate crystal structures and navigate the energy landscape by controlling the coefficients of their amplitudes. Additionally, while we demonstrated the relevance of the use of distortion modes as descriptors through the search for new metastable phases, our method could provide significant insights for many applications. One example is Landau theory for phase transitions \cite{Lifshitz/Landau:1908}
To elaborate, building a Landau model to study phase transitions in materials involves the study of couplings between distortion modes, which is often limited to second or third order \cite{Shapovalov/Stengel:2021} due to the high computational cost. Our model implicitly contains these couplings, which could readily be extracted at minimal computational cost if one would beforehand relax the volume of the structures (in the training set) to avoid the presence of resulting stress contributions to the energy. 
 
Another potential application of our methodology is in the field of ML-driven interatomic potentials with a focus on crystal structure prediction\cite{Deringer/Caro/Csanyi:2019} . 
Constructing machine-learning-guided interatomic potentials involves creating an efficient training set of structures. While a random sampling of the phase space is a valid strategy and has presented great success in elemental systems \cite{Deringer/Pickard/Csanyi:2018}, we believe that our approach can be a useful complement to efficiently exploring phase spaces and predicting energetically accessible polymorphs.
\\

\section*{Acknowledgments}
We thank Andrea Urru and Chiara Gattinoni for helpful discussions and comments on the present work. We also acknowledge financial support from ETH Zürich, the Körber foundation, and the European Research Council (ERC) under the European Union’s Horizon 2020 research and innovation program project HERO grant (No. 810451). Calculations were performed on the Euler cluster at ETH Zürich, and the structures were visualized using VESTA \cite{Momma/Izumi:2011}. 
\bibliography{Bibliography}

\clearpage
\appendix
\renewcommand\thefigure{S\arabic{figure}}    
\setcounter{figure}{0} 
\section*{Appendix}

\begin{figure}[!h]
\includegraphics[width=1\linewidth]{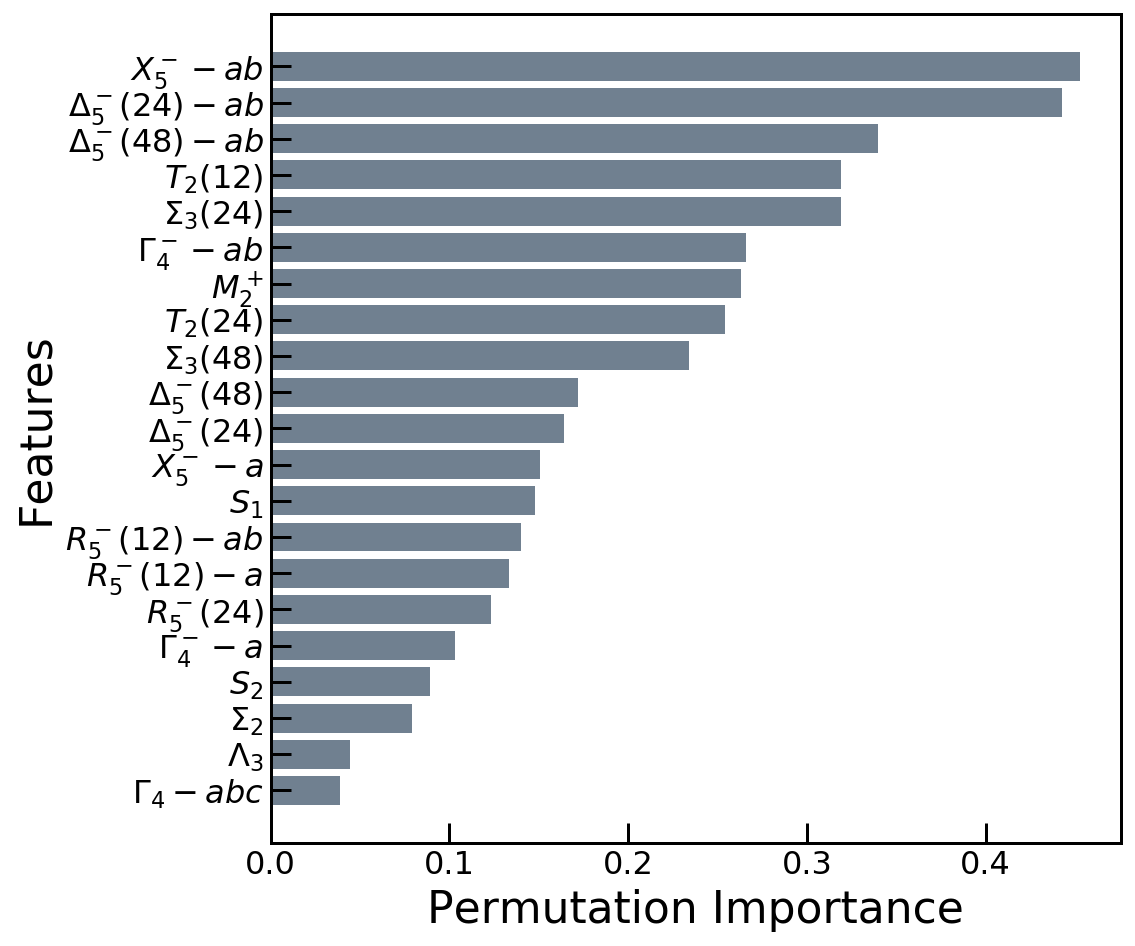}
\centering
\caption{Calculated permutation importance of features of our support vector regression model using 30 repetitions.}
\label{FigS1}
\end{figure}

\begin{figure}[!h]
\includegraphics[width=0.8\linewidth]{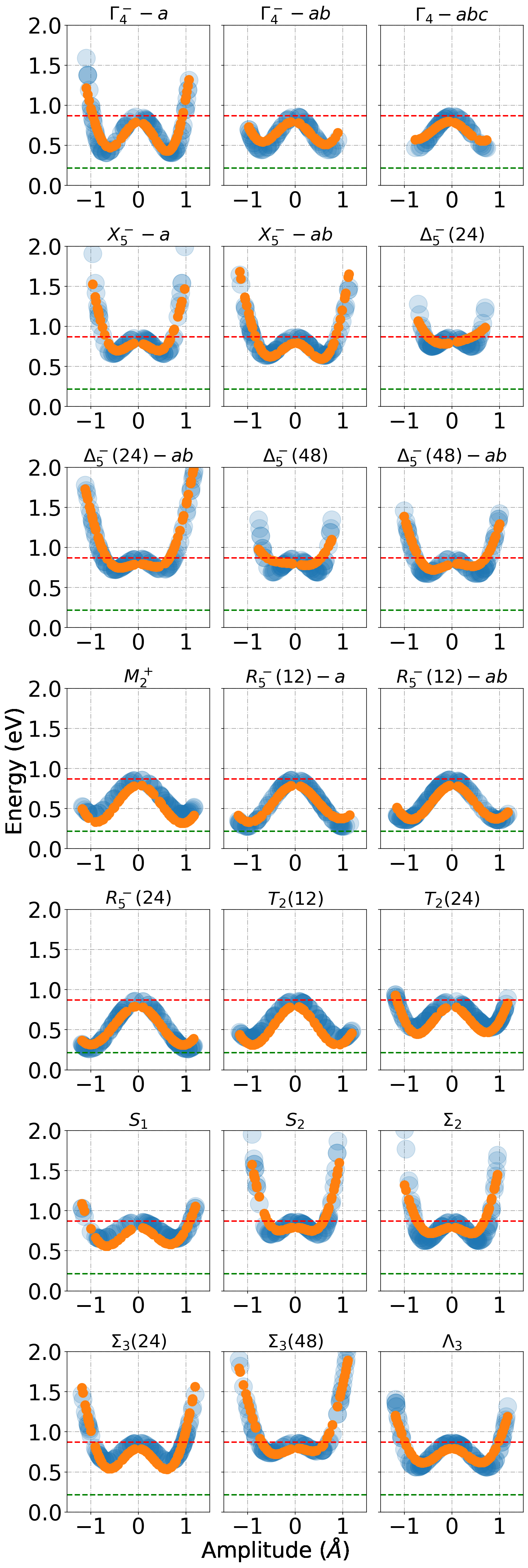}
\centering
\caption{Comparison of the machine-learning predicted variation of the energies as a function of distortion mode amplitude (orange circles) with calculated DFT energies (blue circles).}
\label{FigS2}
\end{figure}

\end{document}